\documentclass{article}
\usepackage{spconf,amsmath,amssymb,graphicx}
\usepackage{cite}
\usepackage{hyperref}
\usepackage{booktabs}
\usepackage[T1]{fontenc}

\newcommand{\subpara}[1]{\vspace{0.5mm}\noindent\textbf{#1}.}

\hypersetup{
    colorlinks=true,
    linkcolor=black,
    citecolor=black,
    filecolor=red,      
    urlcolor=blue,
    pdftitle={d-SynthStrip},
}

\title{Boosting skull-stripping performance for pediatric brain images}

\name{
    \hspace{-5mm}William Kelley$^{1,2}$,
    Nathan Ngo$^{1,2}$,
    Adrian V.\ Dalca$^{1{\text-}3,5}$,
    Bruce Fischl$^{1{\text-}3}$,
    Lilla Z\"ollei$^{1{\text-}3,}$\sthanks{Equal contribution},
    Malte Hoffmann$^{1{\text-}4,\fnsymbol{footnote}}$
}

\address{
    $^1$Athinoula A.\ Martinos Center for Biomedical Imaging, Charlestown, MA 02129, USA \\
    $^2$Department of Radiology, Massachusetts General Hospital, Boston, MA 02114, USA \\
    $^3$Department of Radiology, Harvard Medical School, Boston, MA 02115, USA \\
    $^4$Division of Health Sciences and Technology, MIT, Cambridge, MA 02139, USA \\
    $^5$Computer Science \& Artificial Intelligence Laboratory, MIT, Cambridge, MA 02139, USA \\
}

\begin{document}

\maketitle

\begin{abstract} 
    Skull-stripping is the removal of background and non-brain anatomical features from brain images. While many skull-stripping tools exist, few target pediatric populations. With the emergence of multi-institutional pediatric data acquisition efforts to broaden the understanding of perinatal brain development, it is essential to develop robust and well-tested tools ready for the relevant data processing. However, the broad range of neuroanatomical variation in the developing brain, combined with additional challenges such as high motion levels, as well as shoulder and chest signal in the images, leaves many adult-specific tools ill-suited for pediatric skull-stripping. Building on an existing framework for robust and accurate skull-stripping, we propose developmental SynthStrip (d-SynthStrip), a skull-stripping model tailored to pediatric images. This framework exposes networks to highly variable images synthesized from label maps. Our model substantially outperforms pediatric baselines across scan types and age cohorts. In addition, the <1-minute runtime of our tool compares favorably to the fastest baselines. We distribute our model at \href{https://w3id.org/synthstrip}{https://w3id.org/synthstrip}.
\end{abstract}

\begin{keywords}
    skull-stripping, brain extraction, newborn, infant, toddler, machine learning, pediatric MRI
\end{keywords}

\section{Introduction}
\label{sec:intro}

Skull-stripping is the isolation of the brain from surrounding anatomical features, noise, and background signal in neuroimaging data, for example acquired with magnetic resonance imaging (MRI). It is an essential pre-processing step for many neuroimaging analysis pipelines, in which downstream image processing tasks frequently rely on input images with non-brain tissue removed~\cite{fischl2012freesurfer,zollei2020infant,gaser2022cat}. These pipelines automate labor-intensive processing steps and eliminate subjectivity, enabling researchers to focus on data interpretation and accelerating the pace of discovery in neuroscience.

As neuroanatomy differs substantially between infants and adults, methods developed for the latter are not generally well-suited for younger cohorts. For example, the brain undergoes rapid development during the first two years of life~\cite{knickmeyer2008structural}. During this time the brain doubles in size and the gray-white matter tissue MRI contrast flips (6--9 months). Additionally, pediatric scans are prone to motion artifacts and commonly include parts of the shoulders and chest. These challenges motivate the development of dedicated algorithms for skull-stripping in pediatric populations.

\begin{figure}[b]
     \includegraphics[width=\columnwidth]{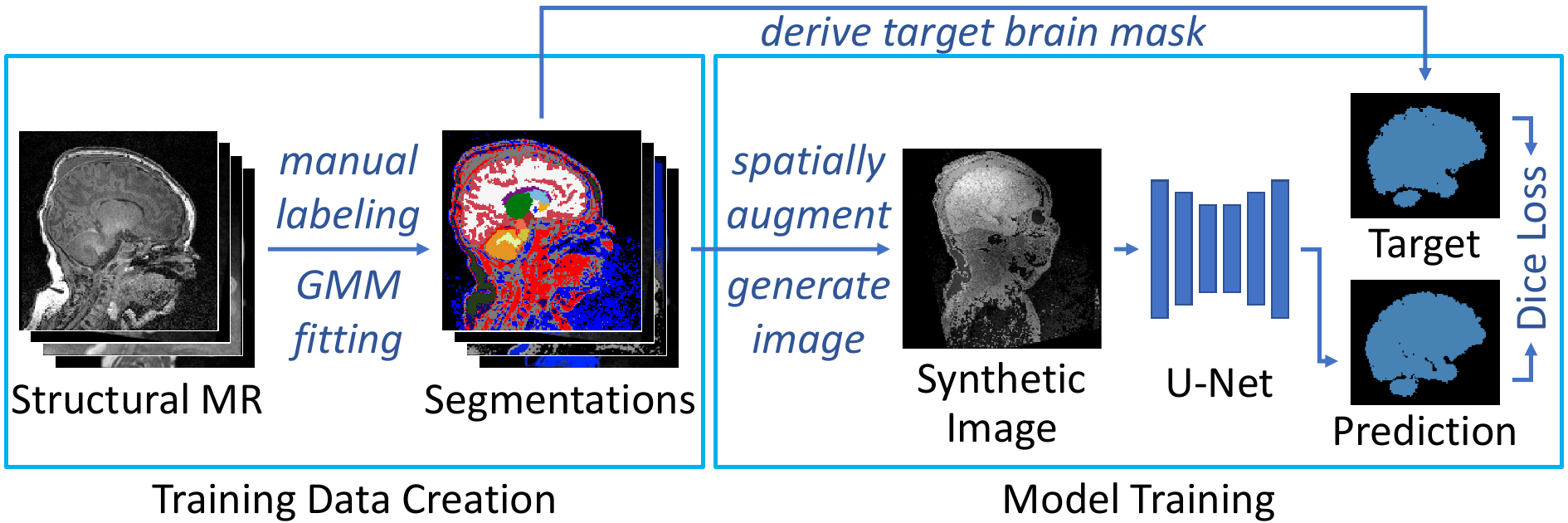}
    \caption{SynthStrip-based training framework. Starting with manual brain label maps, we synthesize widely variable brain images and matching ground-truth brain masks, which we then use to train the model.\label{fig:flowchart}}
\end{figure}

\subpara{Related Work}
There are many existing skull-stripping methods developed for adult brain scans, which leverage a variety of strategies. Some methods iteratively fit deformable brain surfaces to the image~\cite{smith2002fast}, while others determine the brain boundary using a combination of generative and discriminative models, such as Random-Forest classifiers~\cite{iglesias2011robust}. More recently, deep-learning (DL) approaches train deep neural networks to segment the brain~\cite{isensee2019automated}, often building on U-Net architectures~\cite{ronneberger2015u}. 

Few skull-stripping algorithms are tailored specifically to pediatric populations. Typically, these more recent DL methods either target a single MRI contrast~\cite{jog2019fast} or train a different network for each of the available contrasts~\cite{ibeatv20}. One approach uses separate two-dimensional (2D) networks for axial, coronal, and sagittal views extracted from the same input volume before fusing predictions via a voting scheme~\cite{jog2019fast}. Another method trains a 3D U-Net to operate on overlapping 3D patches of the input volume~\cite{ibeatv20}.

\subpara{Synthesis Strategy}
A recent learning strategy trains neural networks without acquired images, producing models that robustly generalize across datasets and imaging modalities~\cite{hoffmann2021learning,billot2023robust}. Synthesizing diverse training images from label maps, prior work achieves state-of-the-art performance on registration~\cite{hoffmann2023anatomy,hoopes2022learning,hoffmann2024joint} and segmentation~\cite{billot2020partial,billot2023synthseg}. SynthStrip~\cite{hoopes2022synthstrip} leverages this approach for robust skull-stripping. Despite demonstrated performance across a large variety of images including pediatric MRI, SynthStrip is an age-agnostic tool that does not specifically target this younger population.

\subpara{Contribution}
We demonstrate that optimizing SynthStrip for pediatric populations leads to performance gains, essential for downstream pediatric neuroimaging pipelines, and helps meet specific pipeline requirements, such as the exclusion of cerebrospinal fluid (CSF) from brain masks~\cite{zollei2020infant}. We build on SynthStrip's generative model and architecture to address the challenges of pediatric neuroimaging data. We create a novel set of pediatric label maps for training-image synthesis and use it to train a new skull-stripping model, developmental SynthStrip (d-SynthStrip). We thoroughly analyze d-SynthStrip's performance on real MRI scans across MRI contrasts and pediatric age groups. We also investigate network-architecture variations to identify an optimal training configuration that surpasses state-of-the-art pediatric solutions in accuracy. Our baseline comparison focuses on publicly available and readily usable tools that can be run without retraining. We will freely distribute our model at \href{https://w3id.org/synthstrip}{w3id.org/synthstrip} as a stand-alone tool and as part of the upcoming FreeSurfer and Infant FreeSurfer releases.

\section{Methods}
\label{sec:method}
We implement the supervised SynthStrip framework~\cite{hoopes2022synthstrip} for skull-stripping and tailor it to pediatric neuroimaging data. Let $x$ be a 3D gray-scale image. A deep convolutional network (CNN) $g_\theta$ with trainable parameters $\theta$ predicts the binary brain mask $\hat{y} = g_\theta(x)$, such that a voxel-wise multiplication yields the skull-stripped image $x_{\hat{y}} = x \odot \hat{y}$.

\begin{figure}
    \includegraphics[width=\columnwidth,trim={0 2mm 2mm 0},clip=true]{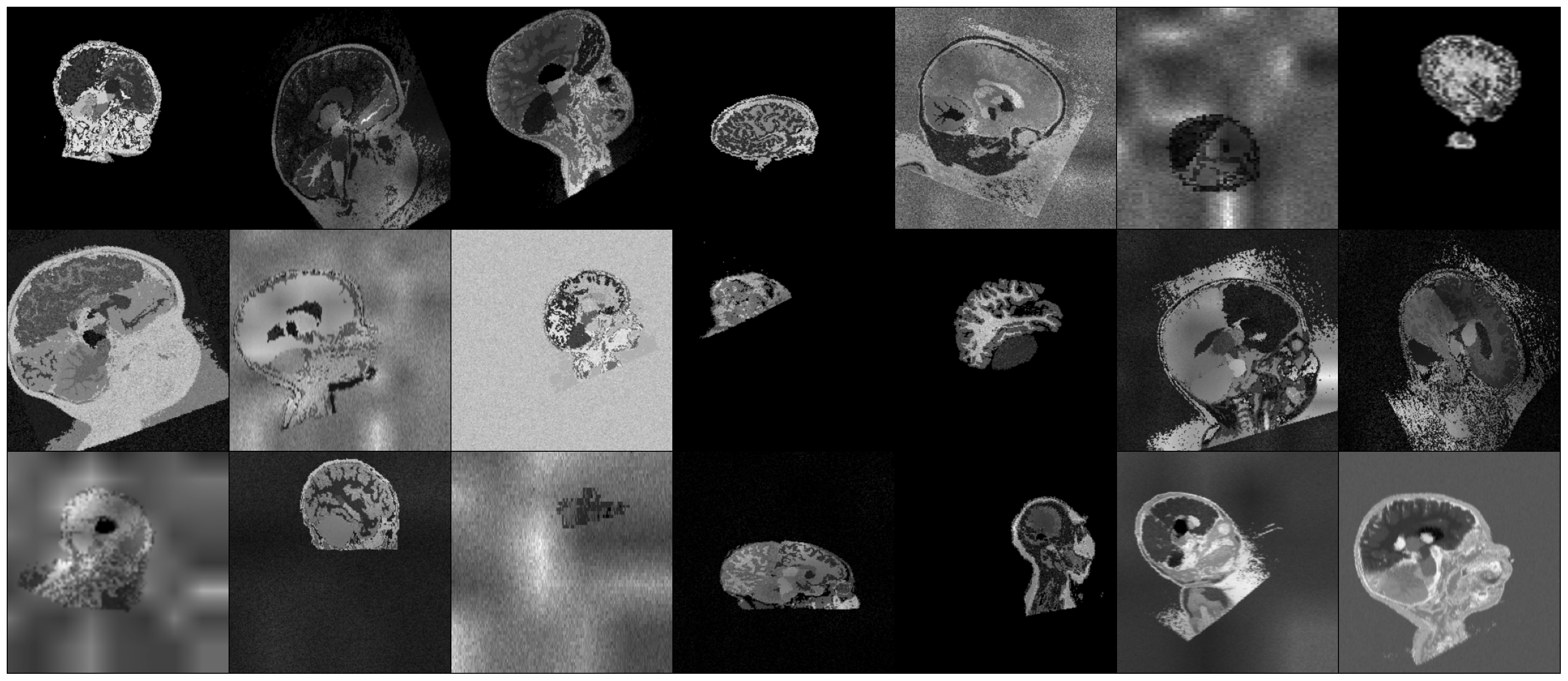} 
    \caption{Synthetic training images generated from pediatric label maps. The spatial and intensity variability deliberately exceeds the range of medical images to encourage d-SynthStrip to generalize across MRI contrasts and age groups.\label{fig:synthedimages}}
\end{figure}

Instead of training with real images, the framework draws a pre-computed whole-head label map $s$ at each optimization step and synthesizes head scan $x$ with randomized intensity features from it. Each step updates parameters $\theta$ to minimize a loss $\mathcal{L}(y, \hat{y})$ that encourages similarity between $\hat{y}$ and the target brain mask $y$, derived from the brain labels in $s$. Figure~\ref{fig:flowchart} provides an overview of the learning framework, while Figure~\ref{fig:synthedimages} shows training-image examples.

\subpara{Training and Validation Data}
We assemble a local dataset (MGH) from (i) 29 Infant FreeSurfer~\cite{zollei2020infant} training images (ii) 18 newborn scans~\cite{warton2018prenatal,warton2021maternal} and (iii) the M-CRIB atlas cohort ($N{=}10$)~\cite{alexander2017mcrib}. We select these 3 sources to cover a wide age range of 0--56 months (Table~\ref{tab:age_data}) and maximize variability across the included structural T1-weighted (T1w) and T2-weighted (T2w) structural scans as well as whole-brain manual label segmentations. We explicitly pool no training subjects from the test datasets (below) to assess generalizability to popular large-scale datasets unseen at training.

We emphasize that we train d-SynthStrip with images synthesized from label maps rather than the label maps themselves. We create training label maps by combining manually drawn brain labels with an additional six labels across the non-brain image content, produced by fitting a Gaussian mixture model (GMM)~\cite{hoopes2022synthstrip} to the intensities of each image. The added labels have no neuroanatomical significance -- we include them in training to synthesize more variable image content. For a balanced distribution of the GMM labels across the image, we apply non-uniformity correction to the image intensities prior to the GMM fit~\cite{fischl2012freesurfer}. For each image, we replace GMM-fitted labels that fall inside the brain boundary with the manual labels to produce a single label map.

\subpara{Generative Model} 
At each training step, we sample $s$ from the set of training label maps~\cite{hoopes2022synthstrip}. First, we augment the spatial variability of $s$ by applying the composition of a random affine (including translation, rotation, scaling, and shear) and nonlinear transform. Second, we sample a mean intensity value for each label and an overall variance. Then we sample intensity values for each voxel of the label from the corresponding normal distribution to generate gray-scale image $x$. Third, we apply an array of randomized corruptions including a spatially-varying intensity bias field, global intensity exponentiation, cropping, downsampling, and Gaussian blurring. These steps produce highly variable training data with complex intensity patterns across the image voxels of each label, including and also far exceeding the variability seen in medical images (Figure~\ref{fig:synthedimages}).

From the spatially augmented label map, we also derive ground-truth brain mask $y$. First, we merge all brain labels excluding non-ventricular CSF to form a binary map. Second, we fill and include the space between brain folds into the brain mask, via 10 iterations of dilation followed by 10 iterations of erosion using nearest-neighbor connectivity. Third, we fill any remaining 3D holes. The resulting brain mask $y$ serves as the target for the network prediction in the loss function.

\subpara{Architecture and Loss}
\label{sec:loss}
We use the 3D SynthStrip U-Net~\cite{hoopes2022synthstrip} architecture. The U-Net $g_\theta$ has seven resolution levels with two leaky-ReLU activated $3\times3\times3$ convolutions per level. It outputs two softmax-activated channels $j$ and $k$ for brain and background, respectively.
We optimize $g_\theta$ using a Dice-based loss $\mathcal{L}_\text{Dice}$, which measures the overlap between the target brain mask $y$ and the predicted mask $\hat{y}$:
\begin{align}
\mathcal{L}_{\text{Dice}}(y, \hat{y}) = -\frac{\sum_{v}{y_j(v) \, \hat{y}_j(v)} + \sum_{v}{y_k(v) \, \hat{y}_k(v)}}{\sum_{v}{y_j(v)^2} + \sum_{v}{\hat{y}_k(v)^2}},
\end{align}
where we sum over all voxels $v \in \Omega$ of the spatial domain $\Omega$ of image $x$.
In our experiments, we also analyze another model variant~\cite{hoopes2022synthstrip}, which predicts a signed distance transform (SDT) $\hat{d}$ representing the distance to the brain boundary at each voxel. We optimize the mean squared error (MSE) from the target SDT $d$ computed from $y$. To focus the optimization gradients on the brain boundary, we down-weight the MSE contribution of voxels farther than distance $h$ from this boundary by a factor of $b$~\cite{hoopes2022synthstrip}.

\subpara{Training Details}
We use 50 label maps from the MGH dataset for synthesis-based training and the remaining 7 real MR images for validation. We train our d-SynthStrip models with stochastic gradient descent using Adam with a batch size of 1, until the loss on the validation set plateaus. We conform all images and label maps to $256^3$ volume size with 1~mm$^3$ isotropic voxels and left-inferior-anterior orientation using linear interpolation.

\section{Experiments}
\label{sec:experiments}

To assess the skull-stripping performance of our models, we compare them against state-of-the-art  baseline methods across MRI contrasts and age groups.

\begin{table}
    \centering
    \newcommand{\same}{\rule[-0.15ex]{0.4pt}{1.7ex}}
    \caption{Age distribution. The dHCP cohort includes preterm and term newborns, listed with gestational age (GA) at scan.\label{tab:age_data}}
    \vspace{2mm}
    \small
    \begin{tabular}{llccccc}
    \toprule
    & & & Min & Max & Mean & St.Dev. \\
    \cmidrule(l){4-7}
    Cohort & Contrast & No. & \multicolumn{4}{c}{Age (months)} \\
    \midrule
    BCP  & T1w & 20 &     5 &    34 &    17 &     8 \\
    BCP  & T2w & 19 & \same & \same & \same & \same \\
    MGH  & \textit{mixed} & 57  & 0 & 56 & 6 & 12 \\
    \midrule
    & & & \multicolumn{4}{c}{GA at scan (weeks)} \\
    \cmidrule(l){4-7}
    dHCP & T1w & 20 &    30 &    43 &    38 &     4 \\
    dHCP & T2w & 20 & \same & \same & \same & \same \\
    \bottomrule
    \end{tabular}
\end{table}

\subpara{Test Data}
\label{sec:testimages}
We select 20 subjects from the UNC/UMN Baby Connectome Project (BCP)~\cite{howell2019unc} and another 20 subjects from the Developing Human Connectome Project (dHCP)~\cite{hughes2017dedicated} to form a test cohort of $N{=}40$ subjects. For each subject, we source a T1w and T2w MR scan along with a label map which corresponds to both images (except 1 BCP subject, for which we have no T2w image). For the BCP cohort, we manually review and correct label maps generated with the Infant FreeSurfer pipeline~\cite{zollei2020infant}. We obtain label maps for the dHCP cohort using the dHCP minimal processing pipeline~\cite{makropoulos2018developing}. Table~\ref{tab:age_data} displays the age distribution for each cohort.

\begin{figure}[b]
    \includegraphics[width=\columnwidth,trim={20mm 14mm 45mm 10mm},clip=true]{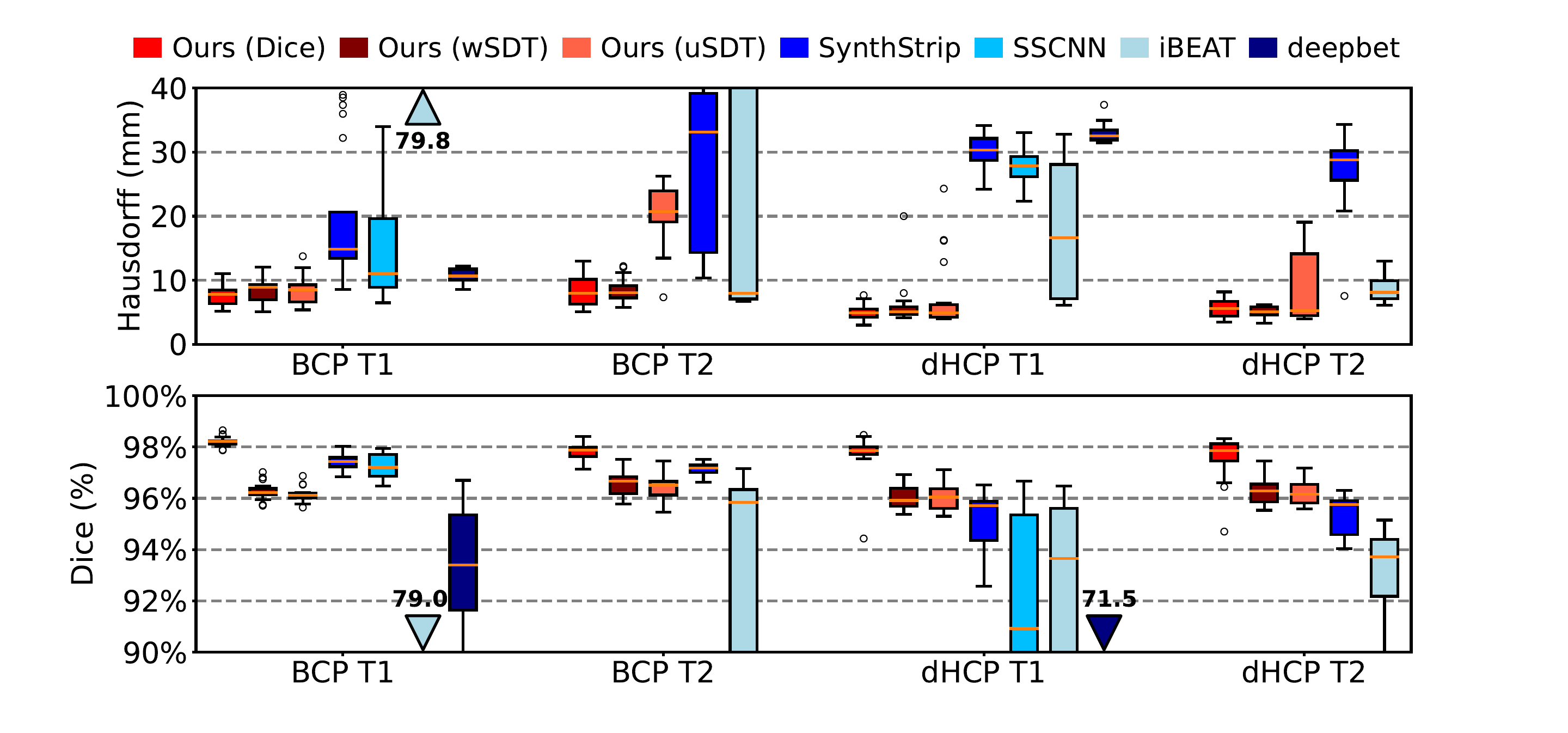} 
    \caption{Brain extraction accuracy in terms of Hausdorff distance and volumetric Dice overlap. Testsets listed in Table~\ref{tab:age_data}.\label{fig:boxplots_combined_smaller_yaxis}}
\end{figure}

\subpara{Baselines}
We compare our tool to well established skull-stripping methods. First, we test SkullStripping CNN (SSCNN)~\cite{jog2019fast}, which targets T1w pediatric MRI. Second, we test the skull-stripping module of the Infant Brain Extraction and Analysis Toolbox (iBEAT)~\cite{ibeatv20} developed for T1w and T2w MRI (version 2.0, release 120). Third, we test SynthStrip~\cite{hoopes2022synthstrip} version 1.5, with the \texttt{--no-csf} flag in order to match the masks predicted by all other methods, which exclude non-ventricular CSF. Finally, we test deepbet~\cite{fisch2023deepbet} version 0.0.2. Although deepbet focuses on T1w adult MRI, we include it as another DL solution due to its demonstrated performance~\cite{fisch2023deepbet}. As deepbet and SSCNN are tailored specifically to T1w MRI, we do not evaluate them on T2w images.

\subpara{Metrics}
We evaluate skull-stripping accuracy relative to binary ground truth masks using volumetric Dice overlap scores and Hausdorff distances between brain-mask boundaries.

\subpara{Setup}
First, we assess the brain-masking accuracy of each tool across MRI contrasts and age groups. Second, we analyze the two different architectures: we compare a traditional segmentation model with a Dice loss to SDT prediction with an unweighted (uSDT, $b=0$~mm) and a weighted SDT loss (wSDT, $b=10^{-3}$, $h=4$~mm) from Section~\ref{sec:loss}.

\begin{figure}
    \includegraphics[width=\columnwidth] {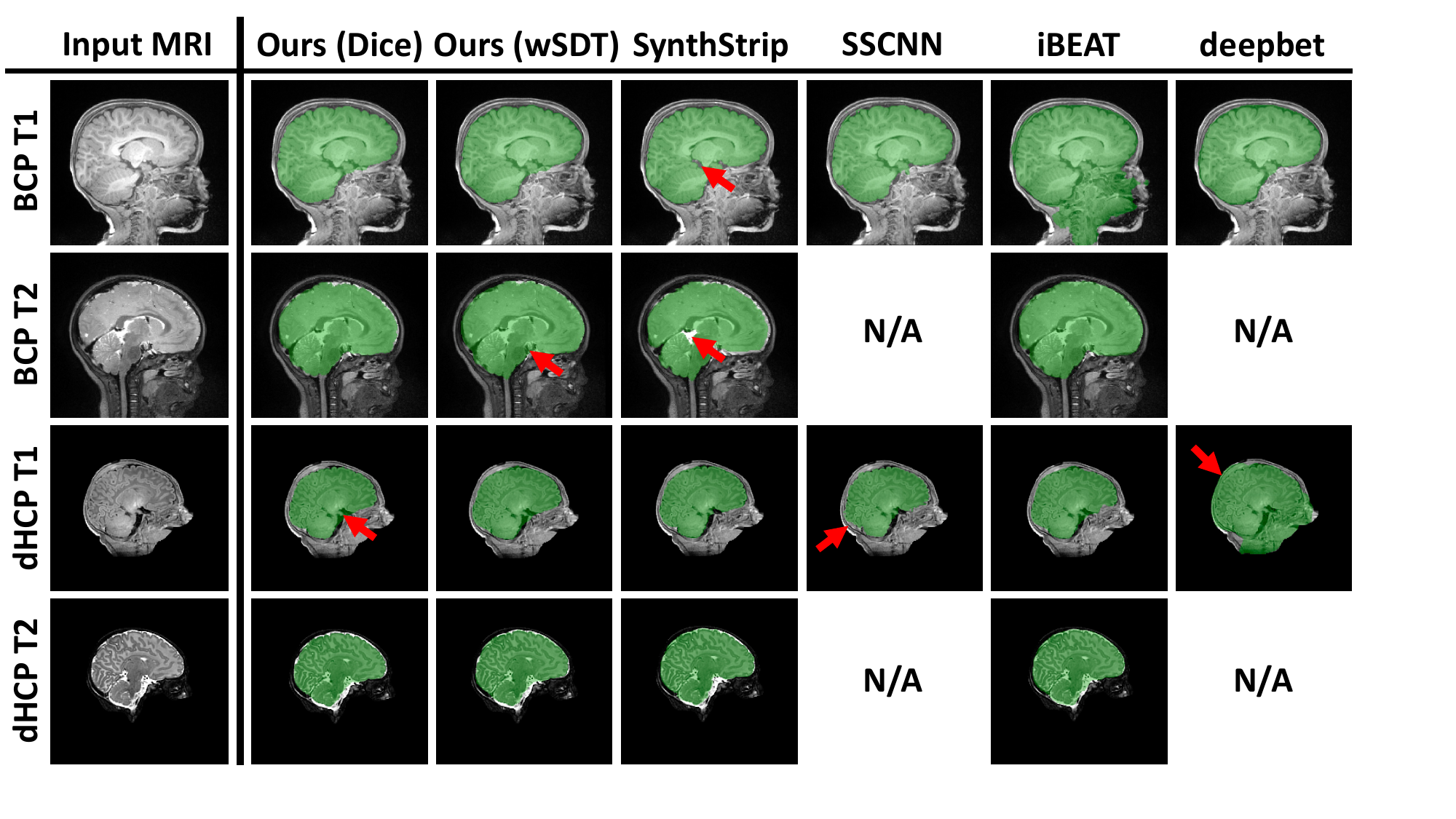} 
    \caption{Representative brain masks predicted by each skull-stripping method. SSCNN and deepbet focus on T1w MRI.\label{fig:examples}}
\end{figure}

\begin{figure}
    \includegraphics[width=\columnwidth]{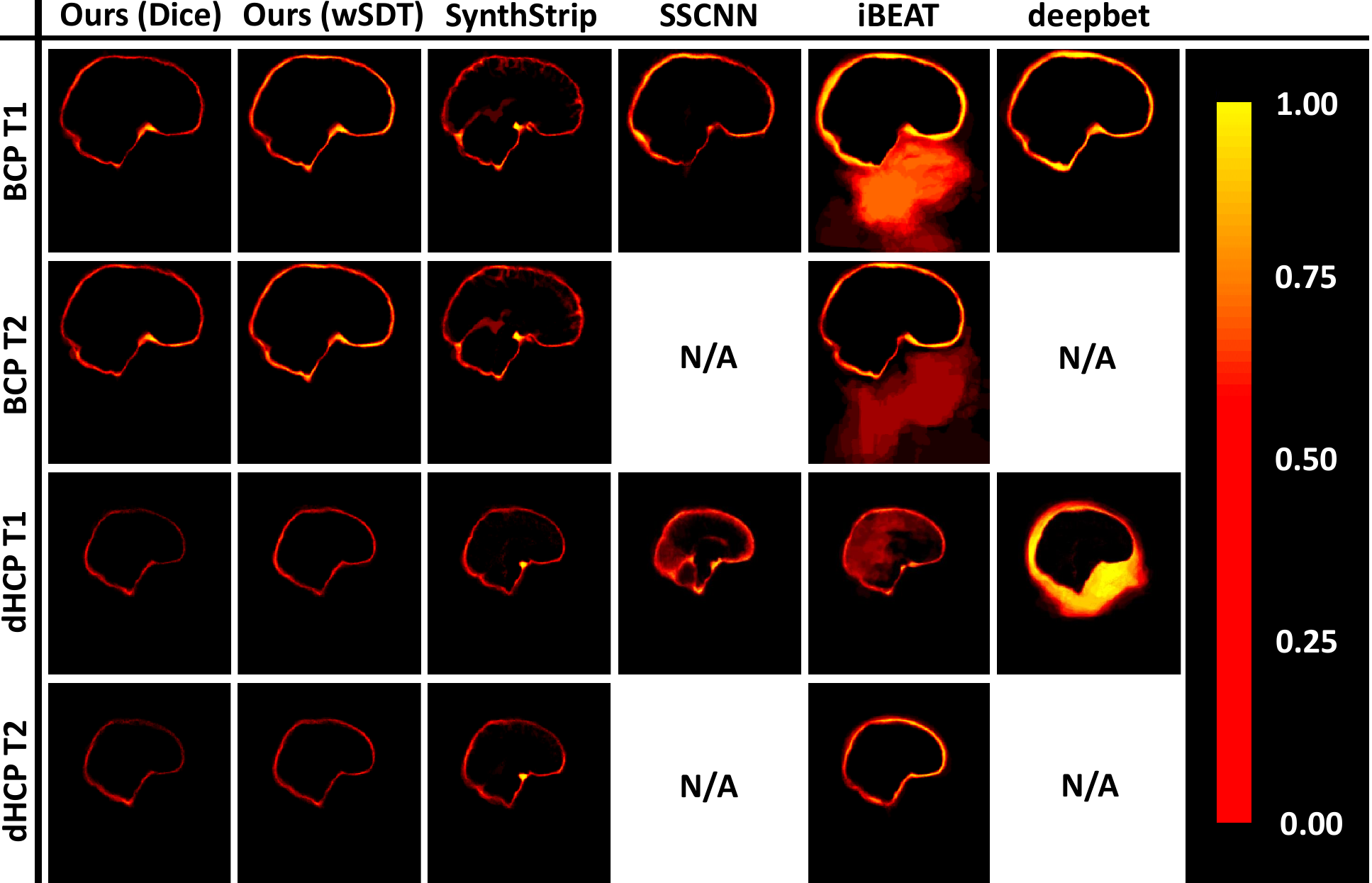}
    \caption{Proportion of absolute skull-stripping errors per voxel in a nonlinear mid-space, across all images of each testset.\label{fig:errors}}
\end{figure}

\subpara{Results} 
\label{ssec:results}
Figure~\ref{fig:boxplots_combined_smaller_yaxis} shows that d-SynthStrip trained with a Dice loss outperforms other skull-stripping methods regardless of contrast or subject cohort. Figure~\ref{fig:examples} compares skull-stripping examples for all methods, and Figure~\ref{fig:errors} quantifies skull-stripping errors across each testset in a nonlinear mid-space. Our SDT models match or slightly under-perform SynthStrip for the BCP images. SSCNN and iBEAT underperform compared to SynthStrip and our model across cohorts except the T1w dHCP scans, where they match the performance of SynthStrip and our d-SynthStrip SDT models.

In terms of Hausdorff distances, both our Dice and SDT models outperform all baselines tested, while the Dice model generally surpasses the SDT variants. SynthStrip closely follows SSCNN. While iBEAT struggles with the BCP data, it achieves the lowest Hausdorff distances among baseline methods for dHCP.

On an NVIDIA RTX 8000 GPU, d-SynthStrip, SynthStrip, and deepbet take less than 1 minute per image, including model setup. However, d-SynthStrip inference alone takes less than 1 second. SSCNN takes approximately 15 minutes, while iBEAT requires up to 22 hours -- skull-stripping results are not available before the full pipeline completes.

\section{Discussion}

We present a pediatric brain extraction tool, d-SynthStrip, that outperforms specialized baseline skull-stripping methods on images acquired from newborns to toddlers. 

While the synthesis strategy previously proved to produce networks that robustly generalize across patient populations, we demonstrate the benefit of synthesizing training data from label maps of a targeted population. d-SynthStrip outperforms SynthStrip by up to 10 Dice points and up to 20~mm Hausdorff distances on infant data. This difference in performance suggests that the synthetic scaling and deformations applied during synthesis may insufficiently cover the distribution of developing brain shapes.

While prior work shows similar skull-stripping accuracy between models trained with Dice and SDT losses~\cite{hoopes2022synthstrip}, we find the Dice loss to lead to increased Dice scores at test time. This result is not surprising, and we plan to investigate receiver operating characteristic (ROC) curves in the future for a more comprehensive comparison of the two losses.

In addition, we will explore whether increasing the variability of the generative model, specifically the synthetic warps applied to input label maps, may bridge the performance gap to yield accurate masks across both pediatric and adult populations with a single model. We will also investigate whether a model trained with a dataset carefully balanced to cover the whole lifespan can robustly accommodate both pediatric and adult brain scans.

\section{Compliance with Ethical Standards}
The retrospective analysis of the MGH data required no ethical approval. We signed data use agreements for access to the publicly available BCP and dHCP data.

\section{Acknowledgments}
The research was supported in part by NIH grants NIBIB P41 EB015896, U01 NS132181, UM1 NS132358, R01 EB023281, R01 EB033773, R21 EB018907, R01 EB019956, P41 EB030006, NICHD R00 HD101553, R01 HD109436, R21 HD106038, R01 HD102616, R01 HD085813, and R01 HD093578, NIA R56 AG064027, R21 AG082082, R01 AG016495, R01 AG070988, NIMH RF1 MH121885, RF1 MH123195, NINDS R01 NS070963, R01 NS083534, R01 NS105820, SIG S10 RR023401, S10 RR019307, as well as S10 RR023043, BICCN U01 MH117023, and Blueprint for Neuroscience Research U01 MH093765. The project also benefited from computational hardware generously provided by the Massachusetts Life Sciences Center.

Data are provided by the dHCP, KCL-Imperial-Oxford Consortium funded by the European Research Council under the European Union Seventh Framework Programme (FP/2007-2013) / ERC Grant Agreement no.\ [319456]. We are grateful to the families who supported this trial.

BF has financial interests in CorticoMetrics, a company focusing on brain imaging and measurement technologies. BF and MH receive salary support from GE HealthCare. Massachusetts General Hospital and Mass General Brigham manage these interests in accordance with their conflict-of-interest policies. The authors have no other interests to disclose.

\bibliographystyle{IEEEbib}
\bibliography{ref}

\end{document}